\documentclass{ws-ijmpcs}
\usepackage{amssymb}

\begin{document}

\markboth{Dai \& Xu} {Quark-cluster Stars: hints from the surface}

%
\catchline{}{}{}{}{}
%

\title{Quark-cluster Stars: hints from the surface}

\author{SHI DAI \and RENXIN XU}

\address{School of Physics and State Key Laboratory of Nuclear Physics and Technology,\\
Peking University, Beijing, 100871, China; \{daishi,
r.x.xu\}@pku.edu.cn}

\maketitle


\begin{abstract}

The matter inside pulsar-like compact stars could be in a
quark-cluster phase since in cold dense matter at a few nuclear
densities ($\rho\sim2-10\rho_{\rm{0}}$), quarks could be coupled
still very strongly and condensate in position space to form quark
clusters. Quark-cluster stars are chromatically confined and could
initially be bare, therefore the surface properties of quark-cluster
stars would be quite different from that of conventional neutron
stars.
Some facts indicate that a bare and self-confined surface of
pulsar-like compact stars might be necessary in order to naturally
understand different observational manifestations. On one hand, as
for explaining the drifting sub-pulse phenomena, the binding energy
of particles on pulsar surface should be high enough to produce
vacuum gaps, which indicates that pulsar's surface might be strongly
self-confined. On the other hand, a bare surface of quark-cluster
star can overcome the baryon contamination problem of $\gamma$-ray
burst as well as promote a successful core-collapse supernova. What
is more, the non-atomic thermal spectra of dead pulsars may indicate
also a bare surface without atmosphere, and the hydro-cyclotron
oscillation of the electron sea above the quark-cluster star surface
could be responsible for those absorption features detected. These
hints could reflect the property of compact star's surface and
possibly the state of condensed matter inside, and then might
finally result in identifying quark-cluster stars.

\end{abstract}

\ccode{PACS numbers: 97.60.Gb, 26.60.Kp, 21.65.Qr}

\section{Introduction}

Pulsar-like compact stars as the densest observable objects in the
universe, are not only important in understanding diverse phenomena
in astronomy, but also significant in fundamental physics.
The answer to the question whether they are neutron or quark stars
would have profound implications on the physics of condensed matter,
the nature of strong interaction as well as the QCD phase
diagram.\cite{Weber,Xu10,Xu11}

Investigating the state of condensed matter inside pulsar-like
compact stars by first principles is extremely difficult. However,
it is meaningful to consider some phenomenological models to
understand observational properties of pulsar-like compact stars. We
conjectured that pulsar-like compact stars could be ``quark-cluster
stars'' which are composed of quark-clusters with almost equal
numbers of up, down and strange quarks, and could be in a solid
state if the kinetic energy of quark clusters (order of $kT$, with
temperature $T$) is lower than the interaction energy between the
clusters.
From a physical point of view, we knew that strong coupling between
quarks is still very strong even in the hot quark-gluon
plasma,\cite{Shuryak2009} so in cold quark matter at realistic
baryon densities ($\rho\sim2-10\rho_{\rm{0}}$), quarks would be
expected to be also coupled strongly and to condensate in position
space, rather than only in momentum space, to form quark clusters.
From an astronomical point of view, the stiff equation of state of
quark-cluster matter could naturally explain the $\sim 2$ solar mass
neutron star,\cite{Lai2011} and the solid state star could provide
gravitational, phase transition and elastic energies to understand
bursts of anomalous X-ray pulsars, soft gamma-ray repeaters and even
the shallow decay of gamma-ray burst afterglow.\cite{Dai2011}

More observational hints come from the surface, since the surface of
quark-cluster stars could be bare and is chromatic confined, which
is quite different from that of conventional neutron stars. One
important surface hint is the
drifting~\cite{Drake,Deshpande,Vivekanand} and
bidrifting~\cite{Qiao} subpulses. The vacuum gap model can explain
drifting subpulses naturally,\cite{RS75} but for conventional
neutron star, the binding energy is too low to form such a vacuum
gap.\cite{Fowlers,Lai} However, for quark-cluster stars, the surface
is chromatic confined and the potential barrier built by the
electric field in the vacuum gap above the polar cap can prevent
electrons from streaming into the magnetosphere, therefore a vacuum
gap can be formed.\cite{Yu} Other important surface hints are
associated with the bare surface of quark-cluster stars. As
discussed in Xu (2002), a quark star could be born to be bare. Such
a bare quark-cluster star can not only provide a clean fireball for
supernovae and gamma-ray bursts, but also explain the thermal
featureless spectrum of neutron stars.\cite{Xu02} What is more,
quark-cluster stars are enveloped in thin electron layers and vortex
hydrodynamical oscillations may be invoked in those electron seas,
which could explain the absorption features detected in the spectra
of some pulsar-like compact stars.~\cite{Xu12}

The outline of the paper is as following. In \S $2$, we briefly
review the quark-cluster star model. In \S $3$, we summarize
different hints from the surface of quark-cluster stars. In \S $4$,
conclusions are presented.

\section{A brief introduction to quark-cluster stars}

Although pulsars have been discovery for more than $40$ years, the
inner structure of pulsar-like compact stars is still in
controversy. Conventional neutron stars and quark stars are two main
types of models. While conventional neutron stars are composed
mainly of hadrons, the equation of supra-nuclear matter should be
understood essentially in the level of quarks.

Previously it is supposed that a perturbative strong interaction
inside quark stars could make quarks to be almost free if the
coupling is weak, whereas for the conventional neutron stars, the
highly non-perturbative strong interaction makes quarks grouped into
neutrons.
However, results of relativistic heavy ion collision experiments
have shown that the interaction between quarks is very strong in hot
quark-gluon plasma.~\cite{Shuryak2009} Therefore, in cold quark
matter at realistic baryon densities ($\rho\sim2-10\rho_{\rm{0}}$),
quarks could also be coupled strongly and the strong interaction
could make quarks to condensate in position space to form quark
clusters.
We may then expect that quark matter inside compact stars could be
in a ``quark clustering phase'', where the energy scale would be
high enough to allow the restoration of 3-light-favor symmetry, but
may not be high enough to make the quarks really deconfined. For a
realistic pulsar-like compact star whose temperature is low, the
kinetic energy of quark clusters should be much lower than the
interaction energy between the clusters, and the stars should in a
solid state.

The peculiar inner structure of quark-cluster stars which are very
different from conventional neutron stars would result in special
global and surface properties. These properties are not only
instructive for us to understand astrophysical phenomena, but also
helpful to identify quark-cluster stars.
Globally speaking, quark-cluster stars have stiff equation of state
with which a compact star could have a maximum mass even to be $>3$
solar mass.~\cite{Lai2011} The heat capacity of quark-cluster matter
is negligible, therefore the cooling behavior would be different
from conventional neutron stars,\cite{YuM} and the missing compact
star of SN1987A could be a quark-cluster star.\cite{LiuXW}
Gravitational and elastic energies of solid quark-cluster stars
could explain bursts of anomalous X-ray pulsars and soft gamma-ray
repeaters;\cite{Tong} the latent heat of phase transition of
quark-cluster matter from liquid to solid could explain the shallow
decay phase of gamma-ray bursts.\cite{Dai2011} As for the surface of
quark-cluster stars, it could be bare and is chromatic confined, and
would be directly associated with the broadband radiation of
pulsar-like compact stars. In the next section, we may discuss the
details about hints from the surface.

In addition, the conjectured quark-cluster matter with 3-flavor
symmetry may also have profound implication to the research of
cosmic rays.
Strangelets composed by quark clusters would exist and propagate in
interstellar space if the gravitational energy is not significant to
form stable quark clusters with strangeness.
It is well know that the most stable hadrons are nucleons (protons
and neutrons), with mass of $\sim 940$ MeV. The simplest particle in
3-flavor symmetry is $\Lambda^0$ ($\sim 1116$ MeV), but the
interaction between $\Lambda$'s is attractive.\cite{LQCD} Therefore
the simplest quark clusters in compact stars would be
$\Lambda$-$\Lambda$ dibaryon, the so-called
$H$-particle.\cite{LGX2011}
A strangelet with mass per baryon $< 940$ MeV (i.e., binding energy
per baryon $\gtrsim  200$ MeV) could be stable in cosmic rays, and
would decay finally into nucleons when collision-induced decrease of
baryon number make it unstable due to the increase of surface
energy.
When a stable strangelet with probably a few (or decades of) quark
clusters bombard the atmosphere of the Earth, its fragmented nuggets
may decay quickly into $\Lambda$'s by strong interaction and further
into nucleons by weak interaction.

\section{Hints from the surface}

Quark-cluster stars could be bare, since proto-quark-cluster stars
may be bare owing to strong mass ejection and high
temperature,\cite{Usov} and accretion of supernova fallback and the
debris disk are unlikely to cover the star with a crust.\cite{Xu02}
For bare quark-cluster stars, surface clusters are confined by
strong colour interaction while the huge potential barrier built by
the electric field can usually prevent electrons from streaming
freely into the magnetosphere.\cite{Yu}

Several observational facts have indicated that such a bare,
self-confined surface could be necessary for us to understand
different phenomena. These hints from the surface of pulsar-like
compact stars may help us identify quark-cluster stars too.

\subsection{Drifting subpulses}

Drifting subpulses were first explained by Ruderman \& Sutherland
(1975). In the seminal paper, a vacuum gap was first suggested above
the polar cap of a pulsar. The sparks produced by the inner-gap
breakdown result in the subpulses, and the $E \times B$ drift which
is due to the lack of charges within the gap, causes the observed
drifting features.
However, the above model encounters the so-called ``binding energy
problem''. Calculations have shown that the binding energy of Fe at
the neutron star surface is $<1$ keV,\cite{Fowlers,Lai} which is not
sufficient to reproduce the vacuum gap. One way to solve this
binding energy problem is the a partially screened inner gap
model,\cite{Gil03,Gil06} but a simple way would be in the bare
quark-cluster star model.\cite{Yu}

In Yu \& Xu (2011), the magnetospheric activity of bare
quark-cluster star was investigated in quantitative details. Since
quarks on the surface are confined by strong color interaction, the
binding energy of quarks can be considered as infinity compared to
electromagnetic interaction. As for electrons on the surface, on one
hand the potential barrier of the vacuum gap prevents electrons from
streaming into the magnetosphere, on the other hand the total energy
of electrons on the Fermi surface is none zero. Therefore, the
binding energy of electrons is determined by the difference between
the height of the potential barrier in the vacuum gap and the total
energy of electrons. Calculations have shown that the huge potential
barrier built by the electric field in the vacuum gap above the
polar cap can usually prevent electrons from streaming into the
magnetosphere, unless the electric potential of a pulsar is
sufficiently lower than that at the infinite interstellar medium.
Figure 1 shows a potential barrier of electrons on the stellar
surface of a typical pulsar.\cite{Yu}

\begin{figure}[th]
\centerline{\psfig{file=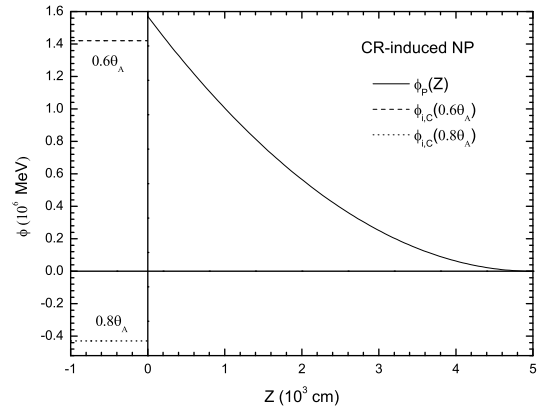,width=8cm}} \vspace*{8pt}
\caption{%
A potential barrier of electrons in the curvature-radiation (CR)
induced vacuum gap of a typical normal pulsar (NP, period $P = 1$ s
and magnetic field $B = 10^{12}$ G). The potential energy of
electrons at the stellar surface, $\phi_{\rm i,C}(\theta)$, is
illustrated with fixed polar angles (e.g. at $0.6\theta_{\rm A}$ and
$0.6\theta_{\rm A}$), where $\theta_{\rm A}$ is the polar angle of
the feet of the last open field lines (Yu \& Xu 2011).
}%
\end{figure}

We conclude that in the bare quark-cluster star model, both positive
and negative particles on the surface are usually bound strongly
enough to form a vacuum gap above its polar cap, and the drifting
subpulses can be understood naturally.

\subsection{A clean fireball for both supernova and $\gamma$-ray burst}

It is well known that the radiation fireballs of gamma-ray bursts
and supernovae as a whole move towards the observer with a high
Lorentz factor.\cite{Paczynski} The bulk Lorentz factor of the
ultrarelativistic fireball of GRBs is estimated to be order of
$\Gamma \sim 10^2-10^3$.\cite{Meszaros} For such an
ultrarelativistic fireball, the total mass of baryons can not be too
high, otherwise baryons would carry out too much energy of the
central engine energy, and this is the so-called ``baryon
contamination''. For conventional neutron stars as the central
engine, the number of baryons loaded with the fireball is unlikely
to be small, since neutron stars are gravity-confined and the
luminosity of fireball is extremely high. However, the baryon
contamination problem can be solved naturally if the central compact
objects are quark-cluster stars. The bare and chromatic confined
surface of quark-cluster stars separate baryonic matter from the
photon and lepton dominated fireball. Inside the star, baryons are
in quark-cluster phase and can not escape due to strong color
interaction, but $e^\pm$-pairs, photons, neutrino pairs and magnetic
fields can escape from the surface. Thus, the surface of
quark-cluster stars automatically generates a low baryon condition
for GRBs as well as supernovae.\cite{Paczynski05,Cheng}

A nascent quark-cluster star born in the center of GRB or supernova
would radiate Planck-like thermal emission due to its ultrahigh
surface temperature,\cite{Haensel} and the photon luminosity is not
constrained by the Eddington limit since the surface of
quark-cluster stars could be bare and is chromatic confined.
Therefore, the strong radiation pressure caused by enormous thermal
emissions from quark-cluster stars might play an important role in
promoting core-collapse supernovae.\cite{Chen} Calculations have
shown that the radiation pressure due to such strong thermal
emission can push the overlying mantle away through photon-electron
scattering with energy as much as $\sim10^{51}$ ergs. Such
photon-driven mechanism in core-collapse supernovae by forming a
quark-cluster star inside the collapsing core is promising to
alleviate the current difficulty in core-collapse supernovae. An
illustration of the photon-driven mechanism is shown in Figure
2.\cite{Chen}

\begin{figure}[th]
\centerline{\psfig{file=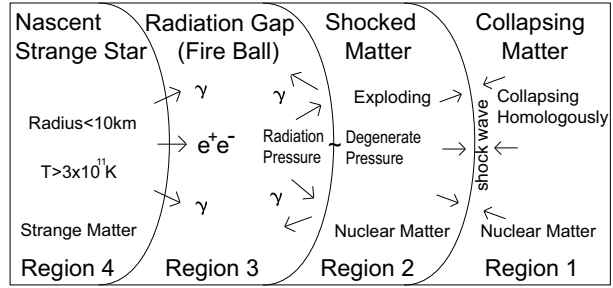,width=8cm}} \vspace*{8pt} \caption{An
illustration of the photon-driven mechanism for core-collapse
supernova (Chen et al. 2007). The outermost region 1 consists of the
unshocked normal matter which is still in-falling, assembled to the
homologous solution. Behind the shock front which serves as the
border and increases in thickness, region 2 comprises the shocked
nuclear matter whose motion has been reversed by the shock. Between
the nascent strange quark-cluster star in the center of the original
collapsing core and region 2 is a fireball (region 3), a gap filled
up with high energy photons and $e^+e^-$ pair plasma, similar to the
fireball of gamma-ray bursts.}
\end{figure}

The nascent quark star could be in a fluid state, but according to
Xu \& Liang (2009), a phase transition of the quark star from liquid
to solid could occur about $10^3-10^6$ s later after its birth.
Since the quark-cluster star in phase transition would emit thermal
radiation at almost constant temperature and a solid quark star
would cool very fast due to its extremely low heat
capacity,\cite{YuM} the latent heat of this phase transition not
only provides a long-term steady central engine, but also shows an
abrupt cutoff of energy injection when the phase transition ends.
Such kind of energy injection may explain the long-lived plateau
followed by an abrupt falloff observed in some
afterglows.\cite{Dai2011}
A schematic cooling behavior of a new-born quark star is shown in
Figure 3.\cite{Dai2011} After the central quark-cluster star is
solidified, star-quake would occur when strains accumulate to a
critical value. The star-quake induced energy ejection would then
results in the observed X-ray flares of bursts.\cite{Xu09}

\begin{figure}[th]
\begin{center}
\centerline{\psfig{file=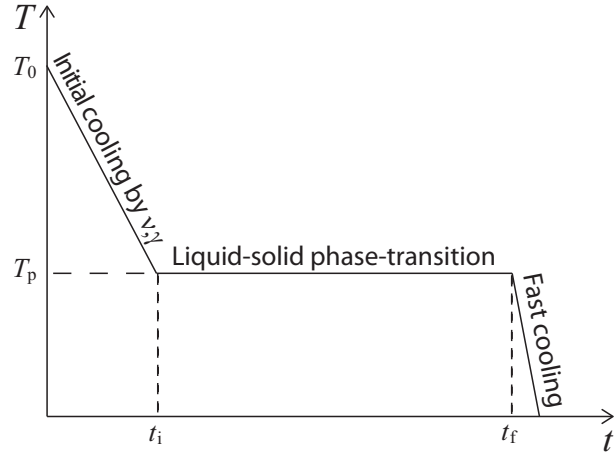,width=8cm}} \vspace*{8pt} \caption{%
A schematic cooling behavior of a new-born quark-cluster star (Dai
et al. 2011). Three stages are shown: an initial cooling stage due
to the emission of neutrinos and photons at the very beginning of a
quark-cluster star with initial temperature $T_{0}$, a liquid to
solid phase transition stage from time $t_{i}$ to $t_{f}$ with
constant temperature $T_{p}$, and, after solidification, a fast
cooling stage because of solid quark star's low heat capacity. The
long-lived plateau followed by an abrupt falloff observed in some
GRB afterglows could be be understood under this scenario.
}%
\end{center}
\end{figure}

\subsection{Non-atomic thermal spectra}

In conventional neutron star/crusted quark stars models, an
atmosphere exists above the surface of central star. The spectrum
determined by the radiative transfer in atmosphere should differ
substantially from Planck-like one, depending on the chemical
composition, magnetic field, etc.\cite{Zavlin} Many calculations
(e.g., Romani (1987); Zavlin, Pavlov, \& Shibanov 1996) show that
spectral lines should be detectable with the spectrographs on board
Chandra and XMM-Newton. One expects to know the chemical
composition, magnetic field of the atmosphere through such
observations, and eventually to determine the mass of the compact
star according to the redshift and pressure broadening of an
absorption spectrum.

However, up to now, no atomic line has been observed with certainty.
Although this discrepancy could be explained for some of the sources
by assuming a low-Z element (hydrogen or helium) photosphere or by
adjusting the magnetic field, we address that the featureless
thermal spectrum can be understood naturally in the bare
quark-cluster star scenario.\cite{Xu02} The best example is the
detailed spectral analysis of the combined X-ray and optical data of
RX J1856.5-3754, which have shown that no atmosphere model can fit
well the data.\cite{Burwitz}

Nevertheless, the best absorption features were detected for the
central compact object (CCO) 1E 1207.4-5209 in the center of
supernova remnant PKS 1209-51/52 at $\sim0.7$ keV and $\sim1.4$
keV.\cite{Sanwal,Mereghetti,Bignami} Although initially these
features were thought to be due to atomic transitions of ionized
helium, soon thereafter it was noted that these lines might be of
electron-cyclotron origin.\cite{Xu03} Recent timing results seems to
agree with the electron-cyclotron
interpretation.\cite{Gotthelf,Halpern}
But this simple single particle approximation might not be reliable
due to high electron density on strange stars, and Xu et al. (2012)
investigated the global motion of the electron seas on the
magnetized surfaces.
It is found that hydrodynamic surface fluctuations of the electron
sea would be greatly affected by the magnetic field, and an analysis
shows that the seas may undergo hydrocyclotron oscillations whose
frequencies are given by $\omega(l)=\omega_{c}/[l(l+1)]$, where
$l=1, 2, 3, ...$ and $\omega_{c}=eB/mc$ is the cyclotron frequency.
The fact that the absorption feature of 1E 1207.4-5209 at $0.7$ keV
is not much stronger than that at $1.4$ keV could be understood in
this hydrocyclotron oscillations model, because these two lines with
$l$ and $l+1$ could have nearly equal intensity, while the strength
of the first harmonic is much smaller than that of the fundamental
in the electron-cyclotron model.
Besides the absorption in 1E 1207.4-5209, the detected lines around
$(17.5, 11.2, 7.5, 5.0)$ keV in the burst spectrum of SGR 1806-20
and those in other dead pulsars (e.g., radio quiet compact objects)
would also be of hydrocyclotron origin.\cite{Xu12}

\section{Conclusions}

The nature of pulsar-like compact stars is closely related to the
physics of condensed matter and the elementary strong interaction.
According to the results of relativistic heavy ion collision
experiments and various properties of pulsar-like compact stars, we
conjecture that quarks could be coupled strongly and condensate in
position space to form quark clusters in cold dense matter at
realistic baryon densities ($\rho\sim2-10\rho_{\rm{0}}$). Under this
scenario, pulsars could be ``quark-cluster stars'' which are
composed of quark-clusters with almost equal numbers of up, down and
strange quarks, and could be in a solid state during almost all of
their lives.

Both global and surface properties of quark-cluster stars are
different from conventional neutron stars. Globally speaking, the
equation of state of quark-cluster stars is stiffer to support
higher mass, while the heat capacity of quark-cluster stars is low,
and solid quark-cluster stars could provide gravitational, phase
transition and elastic energies to various bursts phenomena.
However, more direct hints come from the surface of quark-cluster
stars. Drifting subpulses phenomena may comfortably be understood
with the vacuum gap above the polar cap, and we have noted that such
vacuum gap can easily form above a quark-cluster star surface, while
the binding energy of conventional neutron stars is too low to form
vacuum gap. The bare and chromatic confined surface of quark-cluster
star could naturally overcome the baryon contamination problem in
gamma-ray burst and supernova fireball, and the strong thermal
radiation from quark-cluster star could promote core-collapse
supernovae. The non-atomic thermal spectra of some pulsar-like
compact stars could also be explained by the bare surface of
quark-cluster star, and we have shown that the hydro-cyclotron
oscillation of the electron sea could reproduce observational
frequencies of the absorption features.

All these hints indicate that the surface of pulsar-like compact
star should be bare and self-confined, which is different from
conventional neutron stars. Together with stiff equation of state,
we think that the quark-cluster star model is reasonable to describe
pulsar-like stars. We also expect future astrophysical observations
would identify quark-cluster stars along this line.

Finally, it is worth noting that a high binding ($>200$ MeV per
baryon) between quark clusters not only results in the large maximum
mass of quark-cluster stars,\cite{LGX2011} but also favors one to
conjecture that bulk quark-cluster matter in 3-flavor symmetry could
be absolutely stable.
Compared with nuclear matter (binding energy of $\sim 10$ MeV per
baryon), strange quark-cluster matter is bound strongly, indicating
that the color interaction is still very strong for cold matter at a
few nuclear densities.
No-detection of stellar-mass black holes with mass $<5 M_\odot$ may
hint that the maximum mass for quark-cluster stars is really large.

\section*{Acknowledgments}

We would like to thank useful discussions at our pulsar group of
PKU. This work is supported by the National Natural Science
Foundation of China (Grant Nos. 10935001, 10973002), the National
Basic Research Program of China (Grant Nos. 2009CB824800,
2012CB821800), and the John Templeton Foundation.



\begin{thebibliography}{0}    

\bibitem{Weber}
F. Weber, {\it Prog. Part. Nucl. Phys.}, {\bf 54}, 193 (2005)

\bibitem{Xu10}
R. X. Xu, {\it Int. Jour. Mod. Phys. D}, {\bf 19}, 1437 (2010)

\bibitem{Xu11}
R. X. Xu, {\it Int. Jour. Mod. Phys. E}, to appear (2011)
(arXiv1109.0665)

\bibitem{Shuryak2009}
E. V. Shuryak, {\it Prog. Part Nucl. Phys.}, {\bf 62}, (2009) 48

\bibitem{Lai2011}
X. Y. Lai and R. X. Xu, {\it Research in Astron. Astrophys.}, {\bf
11}, (2011) 687

\bibitem{Dai2011}
S. Dai, L. X. Li and R. X. Xu, {\it Science China G}, {\bf 54},
(2011) 1541

\bibitem{Drake}
F. D. Drake and H. D. Craft, {\it Nat.}{\bf 220}, (1968) 231

\bibitem{Deshpande}
A. A. Deshpande and J. M. Rankin, {\it ApJ}, {\bf 524}, (1999) 1008

\bibitem{Vivekanand}
M. Vivekanand and B. C. Joshi, {\it ApJ}, {\bf 515}, (1999) 398

\bibitem{Qiao}
G. J. Qiao, K. J. Lee, B. Zhang, R. X. Xu and H. G. Wang, {\it ApJ},
{\bf 616}, (2004) 127

\bibitem{RS75}
M. A. Ruderman and P. G. Sutherland, {\it ApJ}, {\bf 196}, (1975) 51

\bibitem{Fowlers}
E. G. Fowlers, J. F. Lee, M. A. Ruderman, P. G. Sutherland, W.
Hillebrandt and E. Muller, {\it ApJ}, {\bf 215}, (1977) 291

\bibitem{Lai}
D. Lai, {\it Rev. Mod. Phys.}, {\bf 73}, (2001) 629

\bibitem{Yu}
J. W. Yu and R. X. Xu, {\it MNRAS}, {\bf 414}, (2011) 489

\bibitem{Xu02}
R. X. Xu, {\it ApJ}, {\bf 570} (2002) 65

\bibitem{Xu12}
R. X. Xu, S. I. Bastrukov, F. Weber, J. W. Yu and I. V. Molodtsova,
2012, {\it Phys. Rev. D}, to appear (arXiv1110.1226)

\bibitem{YuM}
M. Yu and R. X. Xu, {\it Astroparticle Physics}, {\bf 34}, (2011)
493

\bibitem{LiuXW}
X.~W. Liu, J.~D. Liang, R.~X. Xu, J.~L. Han, G.~J. Qiao, (2012)
eprint (arXiv1201.3101)

\bibitem{Tong}
H. Tong and R. X. Xu, {\it Int. Jour. Mod. Phys.}, {\bf E20} (S2),
(2011) 15

\bibitem{LQCD}
T. Inoue et al. {\it Phys. Rev. Lett.}, {\bf 106}, (2011) 162002

\bibitem{LGX2011}
X.~Y. Lai, C.~Y. Gao, R.~X. Xu, eprint (arXiv:1107.0834)

\bibitem{Usov}
V. V. Usov, {\it Phys. Rev. Lett.}, {\bf 80}, (1998) 230

\bibitem{Gil03}
J. Gil, G. Melikidze and U. Geppert, {\it A\&A}, {\bf 407}, (2003)
315

\bibitem{Gil06}
J. Gil, G. Melikidze and B. Zhang, {\it ApJ}, {\bf 650}, (2006) 1048

\bibitem{Paczynski}
B. Paczy\`{n}ski, {\it ApJ}, {\bf 308}, (1986) 43

\bibitem{Meszaros}
P. M\'{e}sz\'{a}ros, M. J. Rees and R. A. M. J. Wijers, {\it ApJ},
{\bf 499}, (1998) 301

\bibitem{Paczynski05}
B. Paczy\`{n}ski and P. Haensel, {\it MNRAS}, {\bf 362}, (2005) 4

\bibitem{Cheng}
K. S. Cheng and Z. G. Dai, {\it Phys. Rev. Lett.}, {\bf 77}, (1996)
1210

\bibitem{Haensel}
P. Haensel, B. Paczy\`{n}ski and P. Amsterdamski, {\it ApJ}, {\bf
375}, (1991) 209

\bibitem{Chen}
A. B. Chen, T. H. Yu and R. X. Xu, {\it ApJ}, {\bf668}, (2007) 55

\bibitem{Xu09}
R. X. Xu and E. W. Liang, {\it Science in China Series G}, {\bf 52},
(2009) 315

\bibitem{Zavlin}
V. E. Zavlin, G. G. Pavlov and Y. A. Shibanov, {\it A\&A}, {\bf
315}, (1996) 141

\bibitem{Romani}
R. W. Romani, {\it ApJ}, {\bf 313}, (1987) 718

\bibitem{Burwitz}
V. Burwitz, V. E. Zavlin, R. Neuh\"{a}user, P. Predehl, J.
Tr\"{u}mper and A. C. Brinkman, {\it A\&A}, {\bf 379}, (2001) 35

\bibitem{Sanwal}
D. Sanwal, G. G. Pavlov, V. E. Zavlin and M. A. Teter, {\it ApJ},
{\bf 574}, (2002) 61

\bibitem{Mereghetti}
S. Mereghetti, A. De Luca, P. Caraveo, W. Becker, R. Mignani and G.
F. Bignami, {\it ApJ}, {\bf 581}, (2002) 1280

\bibitem{Bignami}
G. F. Bignami, P. A. Caraveo, A. De Luca and S. Mereghetti, {\it
Nature}, {\bf 423}, (2003) 725

\bibitem{Xu03}
R. X. Xu, H. G. Wang and G. J. Qiao, {\it Chin. Phys. Lett.}, {\bf
20}, (2003) 314

\bibitem{Gotthelf}
E. V. Gotthelf, J. P. Halpern, {\it ApJ}, {\bf 664}, (2007) 35

\bibitem{Halpern}
J. P. Halpern, E. V. Gotthelf, {\it ApJ}, {\bf 733}, (2011) 28


\end{thebibliography}
\end{document}